\documentclass[12pt]{iopart}

\newcommand{\eqref}[1]{(\ref{#1})}
\usepackage{iopams}  
\usepackage{graphicx}
\usepackage{amsbsy}
\usepackage{color}

\begin{document}

\title[Granular Brownian motion]{Granular Brownian motion}

\author{A. Sarracino, D. Villamaina, G. Costantini and A. Puglisi}

\address{CNR (ISC-SMC) - p.le A. Moro 2, 00185, Roma, Italy}
\address{Dipartimento di Fisica, Universit\`a Sapienza, p.le A. Moro 2, 00185, Roma, Italy}

\ead{ale.sarracino@gmail.com,dario.villamaina@roma1.infn.it}

\begin{abstract}
  We study the stochastic motion of an intruder in a dilute driven
  granular gas.  All particles are coupled to a thermostat,
  representing the external energy source, which is the sum of random
  forces and a viscous drag. The dynamics of the intruder, in
  the large mass limit, is well described by a linear Langevin
  equation, combining the effects of the external bath and of the
  ``granular bath''. The drag and diffusion coefficients are
  calculated under few assumptions, whose validity is well verified
  in numerical simulations. We also discuss the non-equilibrium
  properties of the intruder dynamics, as well as the corrections due
  to finite packing fraction or finite intruder mass.
\end{abstract}

\maketitle

\section{Introduction}

Granular materials in the fluidized state~\cite{JNB96b,PL01} have represented,
during the last 10-15 years, an excellent benchmark for new and old
theories in non-equilibrium statistical mechanics: the
presence of non-conservative forces make unavailable the standard
tools used at equilibrium, such as Gibbs measure, equipartition,
thermodynamic limit, Einstein relation and more~\cite{G99,K99,PBV07,BGM08}. It is
therefore necessary to resort to more fundamental theories,
from the Boltzmann equation up to stochastic processes and modern
generalizations of statistical mechanics to non-equilibrium states~\cite{NE98,BP04}.

In order to achieve a stationary state, the fundamental ingredient is
an external source of energy, required to compensate the energy lost
in inelastic collisions. The role of energy source is played by some
injection mechanisms, depending upon the experimental setup, e.g.: a
box with a vibrating wall, a layer (or more than one) placed over a
vibrating plate, a gas flux going through orifices in the box walls,
etc. The different mechanisms may produce quite different states with
different symmetries: for instance a layer over a vibrating plate is
homogeneous on average, while a boundary driving (e.g.  a shaken box
wall) leads to spatial gradients and currents~\cite{GZB97,WM96,PCV05}.

From the point of view of a tracer particle, however, the dynamics is
always of a similar kind: the tracer interacts, in a random sequence,
with the surrounding particles and with the energy source. The ratio
between frequencies of interaction dictates the relevance of
tracer-particle collisions with respect to exchanges between the
tracer and the source. Of course, in a boundary-driven setup, the
statistics of collisions suffered by the tracer depends upon the
distance from the energy source. Anyway, the random motion performed
by the granular tracer should always take into account the two
contributions: collisions with other granular particles and
interaction with the energy source. In more idealized setups, the
so-called Homogeneous or non-Homogeneous Cooling States, no energy
injection is involved: in this regimes, anyway, a collisional
stationary state cannot be achieved and experimental verification is very
difficult to be achieved.

Here we consider a model commonly used in the theoretical literature
on granular fluids: all grains are coupled to a thermostat-like energy
source, with a typical interaction time $\tau_b$ which is usually
taken larger than the inter-particles collision time
$\tau_c$~\cite{PLMPV98,CDMP04}. The stationary granular gas obtained in this
way, is then used as a ``granular bath'' where a massive intruder
performs a non-equilibrium Brownian motion, still being coupled to
external energy source. The result is a double bath whose properties
are analyzed starting from a linear Boltzmann-Lorentz-Fokker-Planck equation,
which is treated in the diffusional approximation (large mass) to be
cast into a Langevin equation. 

Self-diffusion of an intruder~\cite{BDS99,BRGD99} or a
tracer~\cite{BRCG00,GM04} has been previously studied in the
Homogeneous Cooling State. The same calculations have been performed
for models with an impact velocity dependent restitution
coefficient~\cite{BP05} and a good review of main results can be found
in the textbook~\cite{BP04}. Diffusion in a stationary granular fluid obtained by
imposing shear boundary conditions has also been considered~\cite{GA02}.

Up to our knowledge this is the first time that self-diffusion of a
large mass intruder is studied for a model with homogeneous energy
injection, considering explicitly the effect of a ``double bath'',
i.e. of both sources of noise, granular and external respectively. A
Langevin equation~\eqref{final_langevin} with expressions for the tracer
temperature, Eq.~\eqref{Ttracer}, mobility, Eq.~\eqref{final_drag}, and diffusion
coefficients, Eq.~\eqref{diff}, all involve the interplay of both energy
sources. The large mass limit, together with the Molecular Chaos
assumption (due to diluteness) guarantees that a granular
Fluctuation-Dissipation theorem holds, where the ratio between
diffusion and mobility is simply given by the intruder granular
temperature~\cite{PBL02,DMGBLN03}.

In Section~\ref{themodel} we introduce the model (granular gas,
thermostat and intruder); in Section~\ref{expansion} the Kramers-Moyal
expansion and the large mass limit are discussed, leading to the
Langevin formulation in Section~\ref{langevin}. Numerical experiments (Molecular
Dynamics and Direct Simulation Monte Carlo) are performed to study the limits of the used
assumptions in Section~\ref{simulations} and finally conclusions and
perspectives are drawn in Section~\ref{conclusions}.

\section{The model}\label{themodel}

We consider a gas of $N$ granular spheres in $d$ dimensions, each
sphere has index $i$, with $i\in[1,N]$, and mass $m_i$.
Particle $i=1$ (referred to as ``the intruder'') has mass $M$ and radius
$R$, while all other particles (usually denoted as ``the gas'') have
mass $m$ and radius $r$. The parameter $\epsilon=\sqrt{m/M}$ will be
used for large mass expansion.  The system is contained in a box of
volume ${\cal V}=L^d$, much greater than the volume occupied by the
particles, so that the hypothesis of molecular chaos applies.  We
denote by $n=N/{\cal V}$ the density of the gas and by $\phi$ the
occupied volume fraction (in $d=2$ it is, for instance,
$\phi=\pi[(N-1)r^2+R^2]/{\cal V}$).

The intruder and the gas particles undergo binary instantaneous
inelastic collisions when coming at contact, with the following rule
\begin{eqnarray} \label{colrule}
{\bf v}_i&=&{\bf v}'_i-\frac{m_j}{m_i+m_j}
(1+\alpha)\left[\left({\bf v}'_i-{\bf v}'_j\right)
\cdot\hat{{\bf \sigma}}\right]\hat{{\bf \sigma}} \\
{\bf v}_j&=&{\bf v}'_j+\frac{m_i}{m_i+m_j}
(1+\alpha)\left[\left({\bf v}'_i-{\bf v}'_j\right)
\cdot\hat{{\bf \sigma}}\right]\hat{{\bf \sigma}}, 
\end{eqnarray}
where ${\bf v}_i$ (${\bf v}_j$) and ${\bf v}'_i$ (${\bf v}'_j$) are
the post and pre-collisional velocities of particle $i$ (particle
$j$), respectively; $\alpha\in[0,1]$ is the restitution
coefficient~\footnote{for simplicity we consider the restitution
coefficient to be equal for all particles.}, and $\hat{{\bf \sigma}}$ is the unit
vector joining the centers of the colliding particles. The mean free
path of the intruder  is $l_0=1/(n(r+R)^{d-1})$. Two kinetic temperatures can be
introduced for the two species: the gas granular temperature
$T_g=m\langle v_i^2\rangle/d$ ($i>1$) and the intruder granular temperature
$T_{tr}=M\langle v_1^2\rangle/d$.

In order to maintain a fluidized granular gas, an external energy
source is coupled to every particle in the form of a thermal bath. The
motion of a particle $i$ with velocity ${\bf v}_i$ is then described
by the following stochastic equation
\begin{equation}
m_i\dot{{\bf v}_i}(t)=-\gamma_b {\bf v}_i(t) + {\bf f}_i(t) + \boldsymbol{\xi}_b(t).
\label{langgas}
\end{equation}
Here ${\bf f}_i(t)$ is the force taking into account the collisions with
other particles, and $\boldsymbol{\xi}_b(t)$ is a white noise, with
$\langle\boldsymbol{\xi}_b(t)\rangle=0$ and
$\langle\xi_{b,i\alpha}(t)\xi_{b,j\beta}(t')\rangle=2T_b\gamma_b\delta_{ij}\delta_{\alpha\beta}\delta(t-t')$,
where Latin indices refer to particle labels while Greek indices
denote Cartesian coordinates~\footnote{We use a constant $\gamma_b$,
but in principle this coefficient may depend on the mass and on the radius of the
particle, since it is only a model description of a more complicate
interaction with plates, walls or fluids going through the granular
medium.}.

The effect of the external energy source balances the energy lost in
the collisions and a stationary state is attained. Several temporal
scales are important in this system:
\begin{itemize}

\item $\tau_c^g$, the mean free time between collisions of a gas particle;

\item $\tau_c^{tr}$, the mean free time between collisions of the intruder;

\item $\tau_b^g=m/\gamma_b$ the typical interaction time of the bath with gas particles;

\item $\tau_b^{tr}=M/\gamma_b$ the typical interaction time of the bath with the intruder.

\end{itemize}
When $\gamma_b$ is small enough to have the mean free times $\tau_c^g$
and $\tau_c^{tr}$ smaller than the interaction times $\tau_b^g$ and
$\tau_b^{tr}$, inelasticity is sufficient to put the gas out of
equilibrium: this is reflected, among other things, in the failure of equipartition 
$T_g<T_b$ and $T_{tr}<T_b$. It is also known that $T_g \neq T_{tr}$~\cite{FM02,PMP02}.

The main goal of this note is to show that, in the limit of large mass
$M$, the force ${\bf f}_1$ acting on the intruder can be expressed by
means of a Langevin-like formula
${\bf f}_1(t)=-\gamma_g{\bf V}(t)+\boldsymbol{\xi}_g(t)$, providing explicit
expressions for $\gamma_g$ and $\langle \xi_g(t)\xi_g(t')\rangle$.  

In order to do that, let us start by writing the coupled Boltzmann equations
for the probability distributions $P({\bf V},t)$ and
$p({\bf v},t)$, denoting (for simplicity) with ${\bf V}$ and ${\bf v}$ the
intruder velocity and  the gas velocity, respectively
\begin{eqnarray} \label{beq}
\frac{\partial P({\bf V},t)}{\partial t}&=&\int d{\bf V}'[W_{tr}({\bf V}|{\bf V}')
P({\bf V}',t)-W_{tr}({\bf V}'|{\bf V})P({\bf V},t)] + {\cal B}_{tr}P({\bf V},t) \nonumber \\
\frac{\partial p({\bf v},t)}{\partial t}&=&
\int d{\bf v}'[W_g({\bf v}|{\bf v}')
p({\bf v}',t)-W_g({\bf v}'|{\bf v})p({\bf v},t)] + {\cal B}_g p({\bf v},t) \nonumber \\
&+&J[{\bf v}| p,p],
\label{Boltz}
\end{eqnarray}
where ${\cal B}_{tr}$ and ${\cal B}_g$ are two operators taking into account the interactions with the thermal bath.
In these equations the effects of the collisions for the tracer and the gas particles are described by, respectively,
\begin{eqnarray}
W_{tr}({\bf V}|{\bf V}')&=&\chi\int d{\bf v}'\int d\hat{{\bf \sigma}}
p({\bf v}',t)
\Theta\left[-\left({\bf V}'-{\bf v}'\right)\cdot\hat{{\bf \sigma}}\right]\left({\bf V}'-{\bf v}'\right)\cdot\hat{{\bf \sigma}} \nonumber \\
&\times&\delta^{(d)}\left\{
{\bf V}-{\bf V}'+\frac{\epsilon^2}{1+\epsilon^2}
(1+\alpha)\left[\left({\bf V}'-{\bf v}'\right)
\cdot\hat{{\bf \sigma}}\right]\hat{{\bf \sigma}}\right\}
\label{Wtracer}
\end{eqnarray}
and
\begin{eqnarray}
W_g({\bf v}|{\bf v}')&=&\frac{\chi}{N}\int d{\bf V}'\int d\hat{{\bf \sigma}}
P({\bf V}',t)
\Theta\left[-\left({\bf V}'-{\bf v}'\right)\cdot\hat{{\bf \sigma}}\right]\left({\bf V}'-{\bf v}'\right)\cdot\hat{{\bf \sigma}} \nonumber \\
&\times&\delta^{(d)}\left\{
{\bf v}-{\bf v}'+\frac{1}{1+\epsilon^2}(1+\alpha)\left[\left({\bf v}'-{\bf V}'\right)
\cdot\hat{{\bf \sigma}}\right]\hat{{\bf \sigma}}\right\}, \label{Wgas}
\end{eqnarray}
where $\Theta(x)$ is the Heaviside step function, $\delta^{(d)}(x)$
is the Dirac delta function in $d$ dimensions, and
$\chi=\frac{g_2(r+R)}{l_0}$, $g_2(r+R)$ being the pair correlation
function for a gas particle and an intruder at contact; in the
expressions~\eqref{Wtracer} and~\eqref{Wgas} we have assumed that the
probability $P_2\left(|{\bf x}-{\bf X}|=r+R,{\bf V},{\bf v},t\right)$ that a collision between the
intruder and a gas particle occurs, when they have velocities ${\bf V}$
and ${\bf v}$ and positions ${\bf X}$ and ${\bf x}$ respectively, is given by the Enskog approximation~\cite{BP04}
\begin{equation} \label{enskog}
P_2\left(|{\bf x}-{\bf X}|=r+R,{\bf V},{\bf v},t\right)=g_2(r+R)P({\bf V},t)p({\bf v},t)
\end{equation}
which is a small correction to Molecular Chaos, taking into account density correlations near the intruder;
the terms describing the action of the thermal bath read
\begin{eqnarray}
{\cal B}_{tr}P({\bf V},t)=\frac{\gamma_b}{M}\frac{\partial}{\partial{\bf V}}\left[{\bf V}P({\bf V},t)\right]+
\frac{\gamma_b T_b}{M}\Delta_{V}[P({\bf V},t)]\\
{\cal B}_gp({\bf v},t)=\frac{\gamma_b}{m}\frac{\partial}{\partial{\bf v}}\left[{\bf v}p({\bf v},t)\right]+
\frac{\gamma_b T_b}{m}\Delta_{v}[p({\bf v},t)],
\end{eqnarray}
where $\Delta_v$ is the Laplacian operator with respect to the
velocity; finally, the Boltzmann collision operator for the
particle-particle interactions $J[{\bf v}|p,p]$, can be found in many
papers, see for instance~\cite{GM02}. In view of the fact that it is
not relevant for the rest of the paper, we omit its explicit
expression.

\subsection{Decoupling the gas from the tracer}

The two Boltzmann equations appearing in the system~(\ref{Boltz}) 
are coupled through the terms involving $W_{tr}$ and $W_{g}$. Nevertheless, if the number $N$
of granular particles is large enough, the term $W_{g}$ can be neglected because of the factor $1/N$ in Eq.~(\ref{Wgas}). 
Hence, the surrounding gas is weakly perturbed by the tracer and
fast and homogeneous relaxation is expected.  One
assumes that the probability distribution function $p({\bf v})$ is
stationary and, following numerical evidence (verified below) it is
approximated with a Gaussian function with variance $T_g/m$:
\begin{equation}
p({\bf v})=\frac{1}{\sqrt{(2\pi T_g/m)^d}}\exp\left[-\frac{m{\bf v}^2}{2T_g}\right].
\label{gaussian}
\end{equation}
Substituting Eq.~(\ref{gaussian}) into Eq.~(\ref{Wtracer}), and
projecting the velocities along the collision direction and the
orthogonal one, the integral can be solved~\cite{PVTW06}, yielding
\begin{eqnarray}
W_{tr}({\bf V}'|{\bf V})&=&\chi k(\epsilon)^{-2}(V_{\sigma}'-V_{\sigma})^{2-d}\frac{1}{\sqrt{2\pi T_g/m}}\nonumber \\
&\times&
\exp\left\{-m\left[k(\epsilon)^{-1}\left(V'_{\sigma}-V_{\sigma}\right)+V_{\sigma}\right]^2/(2T_g)\right\},\nonumber \\
\end{eqnarray}
where $V_{\sigma}={\bf V}\cdot\hat{{\bf \sigma}}$ 
(note that $\hat{{\bf \sigma}}$ is parallel to ${\bf V}'-{\bf V}$) and $k(\epsilon)=(1+\alpha)\epsilon^2/(1+\epsilon^2)$.
From now on we specialize to the two dimensional case, where the above
equation simplifies to
\begin{eqnarray}
W_{tr}({\bf V}'|{\bf V})&=&\chi \frac{1}{\sqrt{2\pi T_g/m}k(\epsilon)^2} \nonumber \\
&\times&\exp\left\{-m\left[V'_{\sigma}-V_{\sigma}+k(\epsilon)V_{\sigma}\right]^2/(2T_gk(\epsilon)^2)\right\}.
\label{Wtr2d}
\end{eqnarray}
As discussed in details below, once the gas is decoupled from the intruder,
the dynamics of the tracer alone is Markovian, and it is
known that such transition rates satisfy detailed
balance with respect to a Gaussian invariant probability $P({\bf V})$~\cite{PVTW06}
(the temperature of the tracer, in that case, where $m=M$, is given by
$\frac{\alpha+1}{3-\alpha}T_g$~\cite{MP99}).

\subsection{Granular temperature of the gas}

The granular temperature $T_g$ can be obtained from the Langevin equation~(\ref{langgas}). Indeed,
multiplying by ${\bf v}(t)$ and averaging, one gets
\begin{equation}
\frac{1}{2}m\frac{d}{dt}\langle{\bf v}^2(t)\rangle=-\gamma_b\langle{\bf v}(t)^2\rangle
+\langle{\bf v}(t){\bf f}(t)\rangle+\langle {\bf v}(t)\boldsymbol{\xi}_b(t)\rangle.
\label{langv}
\end{equation}
At stationarity, the l.h.s. of the above equation vanishes and
$\langle {\bf v}(t)\boldsymbol{\xi}_b(t)\rangle=2\gamma_bT_b/m$.  The term
$\langle{\bf v}(t){\bf f}(t)\rangle$ represents the average power
dissipated by collisions, which we assume to be dominated (this is
true for $N$ large enough) by gas-gas collisions:
\begin{eqnarray}
\langle{\bf v}(t){\bf f}(t)\rangle=-\langle \Delta E\rangle_{col},
\label{DeltaE}
\end{eqnarray}
where $\Delta E=1/8m(1-\alpha^2)[({\bf v}_1-{\bf v}_2)\cdot\hat{{\bf \sigma}}]^2$ is the energy dissipated 
per particle and the collision average is defined by
\begin{eqnarray}
\langle\ldots\rangle_{col}&=&\chi_g\int d\hat{{\bf \sigma}}\int d{\bf v}_1\int d{\bf v}_2
\ldots p({\bf v}_1)p({\bf v}_2)\Theta[-({\bf v}_1-{\bf v}_2)\cdot\hat{{\bf \sigma}}]|({\bf v}_1-{\bf v}_2)\cdot\hat{{\bf \sigma}}|.
\nonumber 
\end{eqnarray}
where $\chi_g=\frac{g_2'(2r)}{l_0^g}$ and $l_0^g=1/(n(2r)^{d-1})$ is the mean free path for gas-gas
collisions and $g_2'(2r)$ is the pair correlation function for two gas
particles at contact. The integral in Eq.~(\ref{DeltaE}) can be
computed by standard methods~\cite{BP04}, and, in two dimensions,
yields
\begin{equation}
\langle \Delta
E\rangle_{col}=\chi_g\frac{\sqrt{\pi}(1-\alpha^2)}{\sqrt{m}}T_g^{3/2}.
\end{equation}
Substituting this result into Eq.~(\ref{langv}) and recalling that
$T_g=m\langle {\bf v}^2\rangle/2$, one finally obtains the implicit
equation
\begin{equation}
T_g=T_b-\chi_g\frac{\sqrt{\pi m}(1-\alpha^2)}{2\gamma_b}T_g^{3/2},
\label{Tgas}
\end{equation}
which can be solved to obtain $T_g$.

\section{Kramers-Moyal expansion for the tracer-gas collision operator}\label{expansion}

With the assumption discussed above, the system of
equations~(\ref{Boltz}) is decoupled. That allows us to write the
following linear Master Equation for the tracer
\begin{equation} \label{fp0}
\frac{\partial P({\bf V},t)}{\partial t}=L_{gas}[P({\bf V},t)]+L_{bath}[P({\bf V},t)],
\end{equation}
where $L_{gas}[P({\bf V},t)]$ is a linear operator which can be expressed by means of the Kramers-Moyal
expansion~\cite{R89}
\begin{equation} \label{fpsum}
L_{gas}[P({\bf V},t)]=\sum_{n=1}^\infty \frac{(-1)^n \partial^n}
{\partial V_{j_1}\ldots\partial V_{j_n}}D^{(n)}_{j_1\ldots j_n}({\bf V})P({\bf V},t),
\end{equation}
(the sum over repeated indices is meant) with
\begin{equation} \label{coefficients}
D^{(n)}_{j_1\ldots j_n}({\bf V})=\frac{1}{n!}\int d{\bf V}'(V_{j_1}'-V_{j_1})\ldots(V_{j_n}'-V_{j_n})W_{tr}({\bf V}'|{\bf V}),
\end{equation}
and $W_{tr}$ is given by relation~(\ref{Wtr2d}). The second term in the Master Equation represents the
interaction with thermal bath:  
\begin{equation} \label{bath}
L_{bath}[P({\bf V},t)]={\cal B}_{tr}P({\bf V},t).
\end{equation}
In the limit of large mass $M$, i.e. small $\epsilon$, we expect that the interaction between the granular
gas and the tracer can be described by means of an effective Langevin equation.
In this case, we keep only the first two terms of the expansion~\cite{R89}
\begin{equation} \label{fp}
L_{gas}[P({\bf V},t)]=-\frac{\partial}{\partial V_i}[D_i^{(1)}({\bf V})P({\bf V},t)]+
\frac{\partial^2}{\partial V_i\partial V_j}[D_{ij}^{(2)}({\bf V})P({\bf V},t)].
\end{equation}
A justification of this truncation, in the limit of small $\epsilon$,
comes from observing that terms $D^{(n)}_{j_1 ... j_n}$ are of order
$\epsilon^{2n}$: this can be obtained by plugging
Eqs.~\eqref{colrule} (for the case of the tracer, i.e. ${\bf V}\equiv
{\bf v}_1$) into~\eqref{coefficients}.

It is useful at this point to introduce the velocity-dependent 
collision rate and the total collision frequency
\begin{eqnarray}
r({\bf V})&=&\int d{\bf V}'W_{tr}({\bf V}'|{\bf V}),\\
\omega&=&\int d{\bf V}~P({\bf V})r({\bf V}).
\end{eqnarray}
The former quantity can be exactly calculated, giving
\begin{eqnarray}
r({\bf V})&=&\chi \sqrt{\frac{\pi}{2}}\left(\frac{T_g}{m}\right)^{1/2}e^{-\epsilon^2 q^2/4}
\nonumber \\
&\times&\left[(\epsilon^2 q^2+2)I_0\left(\frac{\epsilon^2 q^2}{4}\right)
+\epsilon^2 q^2 I_1\left(\frac{\epsilon^2 q^2}{4}\right)\right],
\end{eqnarray}
where the rescaled variable ${\bf q}={\bf V}/\sqrt{T_g/M}$ is introduced in Appendix through 
Eqs.~(\ref{rescaled}) and $I_n(x)$ are the modified Bessel functions.
To have an approximation of $\omega$, on the other side, one has to make a position about $P({\bf V})$. 
Let us take it to be a Gaussian with variance $T_{tr}/M$. The consistency of this choice will be verified in the
following section. With this assumption, the collision rate turns out to be
\begin{eqnarray}
\omega=\chi\sqrt{2\pi}\sqrt{T_g/m+T_{tr}/M}=\chi\sqrt{2\pi}\left(\frac{T_g}{m}\right)^{1/2}
\sqrt{1+\frac{T_{tr}}{T_g}\epsilon^2} 
=\omega_0 K(\epsilon) ,
\end{eqnarray}
where $\omega_0=\chi\sqrt{2\pi}\left(\frac{T_g}{m}\right)^{1/2}$ 
and $K(\epsilon)=\sqrt{1+\frac{T_{tr}}{T_g}\epsilon^2}$.

\subsection{Large mass limit}

We are then able to compute the terms $D_i^{(1)}$ and $D_{ij}^{(2)}$ appearing in $L_{gas}$.
The result and the details of the computation of these coefficients as functions of $\epsilon$ 
are given in Appendix. Here, in order to be consistent with the approximation in~(\ref{fp}), 
from Eqs.~(\ref{results}) we report only terms up to ${\cal O}(\epsilon^4)$
\begin{eqnarray} \label{largemass1}
D_x^{(1)}&=&-\chi\sqrt{2\pi}\frac{T_g}{m}q_x(1+\alpha)\epsilon^3+ {\cal O}(\epsilon^5) \nonumber  \\ 
&=&-\chi\sqrt{2\pi}\left(\frac{T_g}{m}\right)^{1/2}(1+\alpha)\epsilon^2 V_x +{\cal O}(\epsilon^5) \nonumber \\
&=&-\omega_0 (1+\alpha)\epsilon^2 V_x +{\cal O}(\epsilon^5) \\
D_y^{(1)}&=&-\omega_0 (1+\alpha)\epsilon^2 V_y +{\cal O}(\epsilon^5) \\
D_{xx}^{(2)}&=&D_{yy}^{(2)}
=\chi\sqrt{\pi/2}\left(\frac{T_g}{m}\right)^{3/2}(1+\alpha)^2\epsilon^4+{\cal O}(\epsilon^5) \nonumber \\
&=&\frac{\omega_0}{2}\frac{T_g}{m}(1+\alpha)^2\epsilon^4 +{\cal O}(\epsilon^5)\\
D_{xy}^{(2)}&=&{\cal O}(\epsilon^6).\label{largemass}
\end{eqnarray}

The linear dependence of $D_\beta^{(1)}$ upon $V_\beta$ (for each
component $\beta$), allows a granular viscosity
\begin{equation}
\eta_g=\omega_0(1+\alpha)\epsilon^2.
\end{equation}
In the elastic limit $\alpha \to 1$, one retrieves the classical
results: $\eta_g \to 2\omega_0 \epsilon^2$ and
$D^{(2)}_{xx}=D^{(2)}_{yy} \to 2\omega_0 \epsilon^2\frac{T_g}{M}$. In
this limit the Fluctuation-Dissipation relation of the second kind is
satisfied~\cite{KTH91,BPRV08}, i.e. the ratio between the noise
amplitude and $\gamma_g$, associated to the same source (collision
with gas particles), is exactly $T_g/M$. When the collisions are
inelastic, $\alpha<1$, one sees two main effects: 1) the time scale
associated to the drag $\tau_g=1/\eta_g$ is modified by a factor
$\frac{1+\alpha}{2}$, i.e. it is weakly influenced by inelasticity; 2)
the Fluctuation-Dissipation relation of the second kind is {\em
violated} by the same factor $\frac{1+\alpha}{2}$. This is only a
partial conclusion, which has to be re-considered in the context of
the full dynamics, including the external bath: this is discussed in
the next section.

\subsection{Langevin equation for the tracer}\label{langevin}

Putting together the results in Eqs.~(\ref{largemass1}-\ref{largemass}) with Eqs.~(\ref{fp0}-\ref{fp}), 
we are finally able to write the Langevin equation for the tracer 
\begin{equation} \label{final_langevin}
M\dot{{\bf V}}=-\Gamma {\bf V}+ {\bf {\cal E}},
\label{langtracer}
\end{equation}
where $\Gamma=\gamma_b+\gamma_g$ and ${\bf {\cal E}}=\boldsymbol{\xi}_b+\boldsymbol{\xi}_g$, with
 \begin{eqnarray} \label{final_drag}
\gamma_g&=&M\eta_g=M\omega_0(1+\alpha)\epsilon^2=\omega_0(1+\alpha)m \label{gammag}\\
\langle {\cal E}_i(t){\cal E}_j(t')\rangle&=& 
2\left[\gamma_b T_b+\gamma_g\left( \frac{1+\alpha}{2}T_g\right)\right]\delta_{ij}\delta(t-t'),
\end{eqnarray}
concluding that the stationary velocity distribution of the intruder is Gaussian with
temperature
\begin{equation} 
T_{tr}=\frac{\gamma_bT_b+\gamma_g\left( \frac{1+\alpha}{2}T_g\right)}{\gamma_b+\gamma_g}.
\label{Ttracer}
\end{equation}
Equation~(\ref{final_langevin}) is consistent with the Gaussian ansatz used in computing
$\omega_0$. Note that the above expression for $T_{tr}$ is consistent
with the large mass expansion obtained in Eqs.~\eqref{largemass} only
if it is dominated by $T_g$, for instance when $\gamma_g \gg \gamma_b$
(see discussion at the end of~\ref{app}). In the opposite
limit, the tracer dynamics is dominated by the coupling with the
external bath and the typical velocity of the tracer cannot be taken
sufficiently small with respect to the typical velocity of gas
particles, making the expansion unreliable. In this case, however, if
the diameter of the intruder is similar to that of the gas particles, it
is reasonable to expect similar collision frequencies: the gas
particles will therefore be dominated by the external bath and the
whole system will be very near to equilibrium~\cite{LCDKG99,PLMPV98}.

For the self-diffusion coefficient it is immediately obtained
\begin{equation} 
D_{tr}=\int_0^{\infty}dt \langle V_x(t)V_x(0)\rangle=\frac{T_{tr}}{\Gamma}
=\frac{\gamma_bT_b+\gamma_g\left( \frac{1+\alpha}{2}T_g\right)}{(\gamma_b+\gamma_g)^2}.
\label{diff}
\end{equation}
Solving numerically the equation~(\ref{Tgas}) and substituting the
result into the above equation, one can study $D_{tr}$ as a function
of the restitution coefficient $\alpha$ (this is done numerically in
the next section). When all other parameters are kept constant and
$\alpha$ is reduced from $1$, the behavior of $D_{tr}$ is
non-monotonic, it decreases, has a minimum and then increases for
lower values of $\alpha$. Anyway, this minimum is expected for quite
low values of $\alpha$ or high values of the packing fraction $\phi$,
where the approximations involved in this theory are not good. For
this reason, at the values of parameters chosen to have a good
comparison with simulations, this non-monotonic behavior is not
observed. 

It should be also noticed that, in the Homogeneous Cooling State, the
self-diffusion coefficient at a given granular temperature increases
as $\alpha$ is reduced from $1$, i.e. it has an opposite behavior with
respect to the present case~\cite{BDS99,BRGD99}. Other studies on
different models of driven granular gases have found
expressions very close to Eq.~\eqref{gammag}, which is not surprising
considering the universality of the main ingredient for this quantity,
i.e. the collision integral~\cite{PBV07,BSL08}.

\subsection{Energy fluxes and detailed balance}

A few comments are in order, at this point, concerning the
non-equilibrium properties of this system. The first question comes
about the term $\frac{1+\alpha}{2}$ which multiplies $T_g$ in
Eq.~\eqref{Ttracer}. It is easily explained with the following
argument~\cite{PVTW06}: we have assumed that the tracer feels no memory of past
collisions, which means that any post-collisional correlation with
recoiling gas particles is lost. With these assumption, the fate of
recoiling particles can be ignored and the dynamics concerns only the
intruder:
\begin{equation}
{\bf V}={\bf V}'-(1+\alpha)\frac{m}{M+m}[({\bf V'}-{\bf v})\cdot \hat{{\bf \sigma}}]\hat{{\bf \sigma}},
\end{equation}
where ${\bf v}$ is the pre-collisional velocity of the colliding gas
particle (randomly extracted from the given distribution $p(\bf{v})$).
Then, one simply observes that for any value of $\alpha$, $M$ and $m$,
such rule can be rewritten as an {\em elastic} collision rule with an
effective mass $M'=2\frac{M+m}{1+\alpha}-m \approx
\frac{2}{1+\alpha}M$ for large intruder mass. This is equivalent to
say that the tracer has elastic interactions with the gas particles,
with an effective mass $M'$, and therefore feels an effective
temperature of the gas $T_g'=\frac{M}{M'}T_g=\frac{1+\alpha}{2}T_g$.
Note that this argument, for $m=M$, gives the formula
$T_g'=\frac{1+\alpha}{3-\alpha}T_g$, which has been derived for the
first time in~\cite{MP99}.

The energy injection rates of the two thermostats~\cite{V06} are
\begin{eqnarray}
Q_b&=\langle {\bf V}(t) \cdot (\boldsymbol{\xi}_b-\gamma_b{\bf V}) \rangle=2\frac{\gamma_b}{M}(T_b-T_{tr})\\
Q_g&=\langle {\bf V}(t) \cdot (\boldsymbol{\xi}_g-\gamma_g{\bf V}) \rangle=2\frac{\gamma_g}{M}(T_g'-T_{tr})
\end{eqnarray}
It is easy to see that the balance of fluxes $Q_b=-Q_g$ is equivalent
to formula~\eqref{Ttracer} for $T_{tr}$. This balance implies that, if
$T_{tr}<T_b$, then $T_{tr}>T_g'$. When $\alpha<1$, the two fluxes are
different from zero, i.e. energy is flowing from the external driving,
through the tracer, into the granular bath.

Apparently, this contradicts the ``equilibrium'' nature of the
Langevin equation~\eqref{final_langevin}: the tracer dynamics is
Markovian and stationary, and the equation satisfies detailed
balance with respect to the Gaussian invariant distribution.  As
already discussed in~\cite{PVTW06}, this is not a paradox but only a
consequence of Molecular Chaos and the decoupling assumption
which allows us to write Equation~\eqref{fp0}: here we have employed the
Enskog approximation, which is a weak modification of Molecular Chaos,
still preserving Markovianity, i.e. no memory terms appear in
Eq.~\eqref{beq}. The absence of memory implies that both
$\xi_b$ and $\xi_g$ are white noises and makes them indistinguishable: an
observer which can only measure ${\bf V}(t)$ cannot obtain separate
measures of $Q_b$ and $Q_g$, but only a measure of the total energy
flow $Q=M\langle {\bf V} \cdot \dot{{\bf V}}\rangle=0$ which hides out
the presence of energy currents. A more detailed analysis, e.g. by
relaxing the Enskog approximation, should put in evidence the
different time-correlations of the two baths: eventually, the
observer, by means of some ``filter'', should be able to sort
out their different contributions $Q_b$ and $Q_g$. This is an interesting
example where memory plays a crucial role in the non-equilibrium
characterization of a system~\cite{PV09}.

We expect that time reversibility (detailed balance) is a symmetry,
for the intruder, which is broken in the following cases: 1) at small
values of $M$ (this is different from the case discussed
in~\cite{PVTW06}, where the intruder was not in contact with the
external bath); 2) when the non-Gaussian behavior of the gas
velocities is taken into account; 3) when the tracer has asymmetric
properties with respect to some spatial axis~\cite{CPB07}; 4) when
Molecular Chaos (or its weak Enskog correction) is violated~\cite{PBV07}.

\section{Numerical simulations}\label{simulations}

In this Section we report the results of Molecular Dynamics (MD)
simulations of the model, together with Direct Simulation Monte Carlo
(DSMC) simulations~\cite{B94} incorporating the Enskog correction, and
compare them with our theoretical predictions. In all simulations we
have kept constant the dimension $d=2$, the mass of gas particles
$m=1$ and the radii $r=R=0.01$, as well as the properties of the bath
$T_b=1$ and $\gamma_b=0.1$; instead we have varied $N$, $M$, $\alpha$
and $\phi$ (values of $L$ and $n$ can be obtained from the knowledge
of $r$ and $\phi$). We have used the Carnahan-Starling expression for
$g_2$ at contact~\cite{CS69}: $g_2(r+R)=(1-\frac{7}{16}\phi)/(1-\phi)^2$.
For the chosen values of $\phi \leq 0.07$, it is always $g_2(r+R) \leq
1.12$.  In all simulations we have also checked that the Gaussian
approximations for the velocity distributions of gas particles and for
the intruder are satisfied, observing very small values for the second
Sonine coefficient $a_2\leq0.02$~\cite{BP04}.

\begin{figure}[!htb]
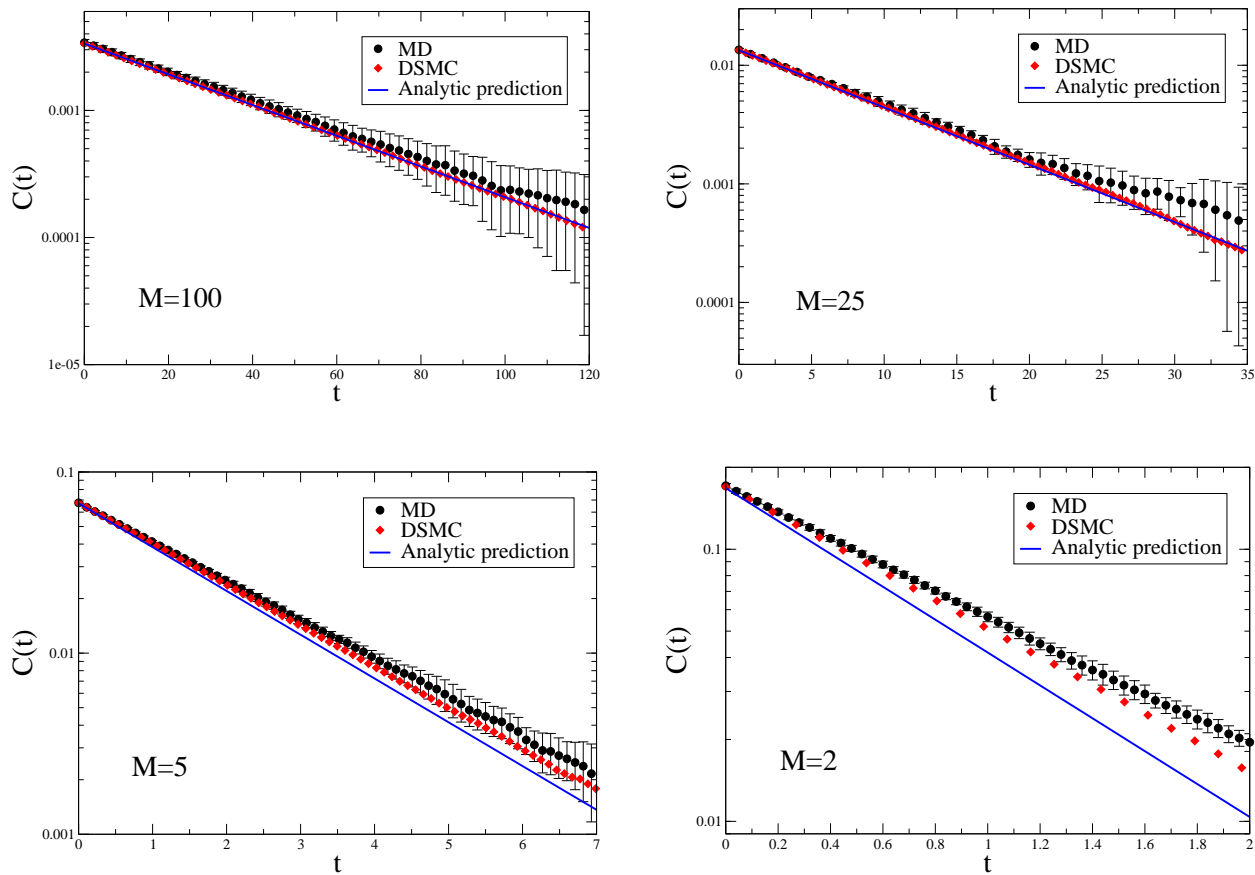


\includegraphics[width=0.5\columnwidth,clip=true]{M100.eps}\qquad\qquad
\hspace*{-1.0cm}
\includegraphics[width=0.5\columnwidth,clip=true]{M25.eps}

\vspace*{0.8cm}

\includegraphics[width=0.5\columnwidth,clip=true]{M5.eps}\qquad\qquad
\hspace*{-1.0cm}
\includegraphics[width=0.5\columnwidth,clip=true]{M2.eps}

\caption{(Color online). The autocorrelation function $C(t)=\langle V_x(t)V_x(0)\rangle$ 
is measured in MD and DSMC simulations (black circles and red diamonds, respectively)
for $M=100,25,5,2$ in the model with restitution coefficient $\alpha=0.8$
and packing fraction $\phi=0.00785$ and coupled to a thermal 
bath with $\gamma_b=0.1$ and $T_b=1$. The blue lines show the theoretical
predictions of Eq.~(\ref{cdit}).}
\label{varieM}
\vspace{1cm}
\end{figure}

In Fig.~\ref{varieM} we show the velocity-velocity autocorrelation
function $C(t)=\langle V_x(t)V_x(0)\rangle$ of the tracer for
different values of its mass $M=100,25,5,2$ in a dilute and moderately
inelastic case: $\alpha=0.8$, and $\phi=0.00785$ (and $N=10^4$ for MD). We can
clearly observe that in the case of large mass $M=100$ the Langevin
equation~(\ref{langtracer}) describes very well the dynamics of the
tracer. Indeed, in that case, the numerical results are consistent
with the theoretical prediction
\begin{equation}
C(t)=\frac{T_{tr}}{M}e^{-\frac{\Gamma}{M}t}.
\label{cdit}
\end{equation}
As expected, for smaller values of $M$, the numerical results move
away from the analytical ones and large corrections to the exponential
decay do appear. The deviations are observed (and are quantitatively
similar) for both MD and DSMC results, implying that they are due, as
expected, to the breakdown of the large mass expansion, rather than that of
Molecular Chaos. For MD results we have noticed that, going from
$N=10^3$ to $N=10^4$, the comparison with DSMC (and with theory at
large $M$) is improved. 

\begin{figure}[!htb]
\begin{center}

\includegraphics[width=0.5\columnwidth,clip=true]{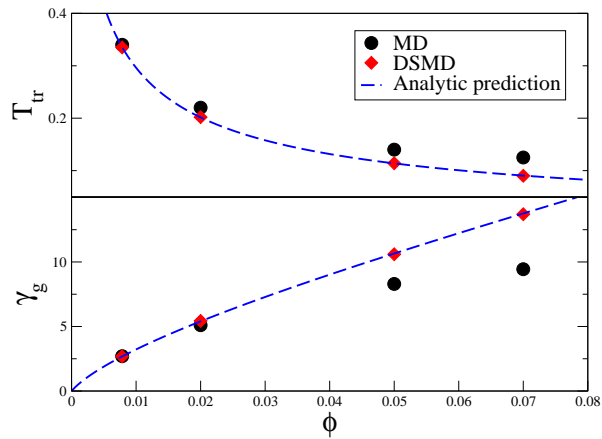}

\caption{(Color online). The temperature $T_{tr}$ (top panel) and the
drag coefficient $\gamma_g$ (bottom panel) measured in MD (black circles)
and DSMC (red diamonds) is plotted for 
different values of the packing fraction $\phi=0.00785,0.2,0.5,0.7$ in the model with $M=100$,
$N=10^4$ (in MD) and  $\alpha=0.8$ (error bars fall within the symbols). 
The dashed blue lines show the theoretical
predictions following from Eqs.~(\ref{Ttracer},\ref{gammag}).}
\label{fig:variePhi}
\end{center}
\end{figure}

In order to check the validity of the hypothesis of molecular chaos,
we report the results of MD and DSMC simulations for higher packing
fractions in Fig.~\ref{fig:variePhi}, keeping $M=100$, $N=10^4$ (in
MD) and $\alpha=0.8$: since the clean part of the decay of $C(t)$ is
always exponential, we focus only on the two parameters of interest,
i.e. $T_{tr}$ and $\gamma_g$.  One clearly observes that, increasing
the packing fraction, the discrepancy between the theoretical value
and the values obtained from MD, increases. On the other side, DSMC
always gives results very close to theory, as expected. The Enskog
approximation~\eqref{enskog}, which does not take into account memory
effects, is no longer valid in MD at high packing fraction, while
always holds in DSMC.  In order to enforce this statement, we computed
the following correlation coefficient: $C_{V u_{m}}=\frac{\left<\delta
V_x \delta u_{m} \right>}{\sqrt{\langle \delta V_x^2 \rangle}\sqrt{\langle \delta u^2_{m} \rangle}}$,
where we introduced the stochastic variable $u_{m}(t)$ given by the
averaged $x$-component velocity of the particles lying, at time t, in
a fixed area around the tracer. In particular, $\delta V_x$ and
$\delta u_{m}$ measure the deviations of $V_x$ and $u_{m}$ from the
average values, which tend to $0$ for a large number of measures.  The
coefficient defined above must be zero, if molecular chaos holds; on
the contrary we observed that its value sensibly increases as the
packing fraction gets higher. For example for $\phi=0.00785$
,$C_{Vu_{m}}=0.005$ whereas, for $\phi=0.07$, $C_{Vu_{m}}=0.07$.

\begin{figure}[!htb]
\begin{center}

\includegraphics[width=0.5\columnwidth,clip=true]{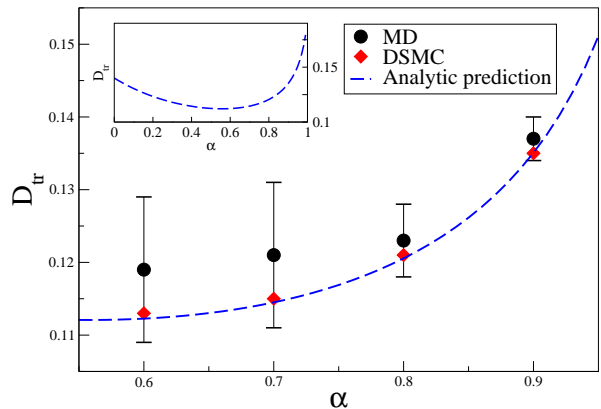}

\caption{(Color online). The diffusion coefficient of the tracer
$D_{tr}$ is measured in MD (black circles) and DSMC (red diamonds)
simulations for different values of the restitution coefficient
$\alpha=0.9,0.8,0.7,0.6$ in the model with $M=100$ and packing
fraction $\phi=0.00785$. The dashed blue line shows the theoretical
prediction following from Eqs.~(\ref{Tgas},\ref{diff}). In the inset
the same curve is plotted in the whole range $\alpha \in [0,1]$.}
\label{variAlpha}
\end{center}
\end{figure}

Finally let us compare the diffusion coefficient
$D_{tr}=\int_0^{\infty}dt~C(t)$ measured in MD and DSMC with the
theoretical value obtained through Eqs.~(\ref{Tgas}) and~(\ref{diff}).
In Fig.~\ref{variAlpha} we show our results at different values of
$\alpha$, keeping fixed $M=100$, $\phi=0.00785$ and $N=10^4$ (in
MD). Again there is a perfect match for DSMC, while MD simulations
present a small discrepancy which becomes more evident at small values
of $\alpha$. We have again verified that this discrepancy is a finite
$N$ effect and is reduced as $N$ increases.

\section{Conclusions}\label{conclusions}

While many papers have been devoted to the large-mass diffusive
properties of an intruder in a cooling granular gas, the driven case,
somehow, has received less attention~\cite{GA02}: this is in contrast with the
fact that, in real experiments, the most common situation is a driven
granular gas. The problem, at the level of the basic assumptions
treated here (Enskog approximation, negligible non-Gaussianity and
large separation of timescales between collisions and driving), does
not pose particular conceptual difficulties, nevertheless it reveals to be
already quite rich. The external driving mechanism, characterized by a
temperature $T_b$ and the ``internal'' granular bath at temperature
$T_g<T_b$, sum up together in giving a linear Langevin dynamics for
the intruder, provided that the collision frequency between the
intruder and the gas particles is larger than the frequency of
interaction with the bath. Such Langevin equation predicts for the
``intruder temperature'' $T_{tr}$ a weighted sum (with weights given
by the drag coefficients of the two baths) of $T_b$ and
$T_g'=\frac{1+\alpha}{2}T_g$, i.e. the intruder feels the surrounding
gas to be at a different temperature $T_g'<T_g$, because of
non-conservative interactions. The self-diffusion coefficient is even
more interesting, showing a non-trivial non-monotonic behavior with a
minimum at low values of the restitution coefficient. Our results lose
validity when the mass of the intruder is reduced, when the packing
fraction of the gas is increased, when the inelasticity is too low to
disregard non-Gaussian corrections, and when the interaction times of
the two baths become comparable.

It is interesting to discuss what is happening at moderately high
packing fractions $\phi\sim 10\%$: we have seen that the Enskog
approximation is not very good to predict the intruder dynamics,
because it is missing memory effects mediated by the surrounding
fluid. A scenario which can be conjectured is the following: the gas
may display two typical relaxation times, a local one related to
collisions $\tau_{rel}\sim \tau_c^g$ and a global one
$\tau_{rel}'>\tau_{rel}$, which is due to diffusion of slower modes
(e.g. hydrodynamics). If $\tau_{rel}'>\tau_c^{tr}>\tau_{rel}$, one has
that the intruder feels a ``locally equilibrated'' surrounding
granular gas. In this case it is reasonable to replace
Eq.~\eqref{gaussian} with
\begin{equation}
p({\bf v})=\frac{1}{\sqrt{(2\pi T_g/m)^d}}\exp\left[-\frac{m({\bf v}-{\bf u})^2}{2T_g}\right]
\label{localgaussian}
\end{equation}
where ${\bf u}$ and $T_g$ are some local velocity and temperature fields
which change on timescales larger than $\tau_c^{tr}$ (and
correspondingly large spatial scales). A partial verification of this
scenario has been mentioned at the end of~\cite{PBV07}, but requires further investigation.

\ack The work of all authors is supported by the ``Granular-Chaos''
project, funded by the Italian MIUR under the FIRB-IDEAS grant number
RBID08Z9JE. The authors are also indebted with Paolo Visco and Angelo
Vulpiani for useful discussions and a careful reading of the
manuscript.

\appendix
\section{Calculation of first two coefficients of the Kramers-Moyal expansion}
\label{app}

For larger generality (whose motivation is discussed in the
Conclusions), in this Appendix we discuss the case where the gas
surrounding the intruder may have a non-zero average ${\bf u}$~\footnote{note
that in all the cases discussed in the main text, we have always taken
${\bf u}=0$.}:
\begin{equation}
p({\bf v})=\frac{1}{\sqrt{(2\pi T_g/m)^d}}\exp\left[-\frac{m({\bf v}-{\bf u})^2}{2T_g}\right]
\end{equation}
which is a simple task involving only the definition of new shifted variables
\begin{eqnarray}
{\bf c}&=&{\bf V}-{\bf u}\\
{\bf c}'&=&{\bf V}'-{\bf u}.
\end{eqnarray}
We are interested in computing
\begin{eqnarray}
D_i^{(1)}({\bf V})&=&\int d{\bf V}'(V_i'-V_i)W_{tr}({\bf V}'|{\bf V}) \nonumber \\
&=&\int d{\bf c}'(c_i'-c_i)\chi\frac{1}{\sqrt{2\pi T_g/m}k(\epsilon)^2} \nonumber \\
&\times&\exp\left\{-m\left[c'_{\sigma}+(k(\epsilon)-1)c_{\sigma}\right]^2/(2T_gk(\epsilon)^2)\right\}. 
\end{eqnarray}

\begin{figure}
\includegraphics[clip=true,width=7cm]{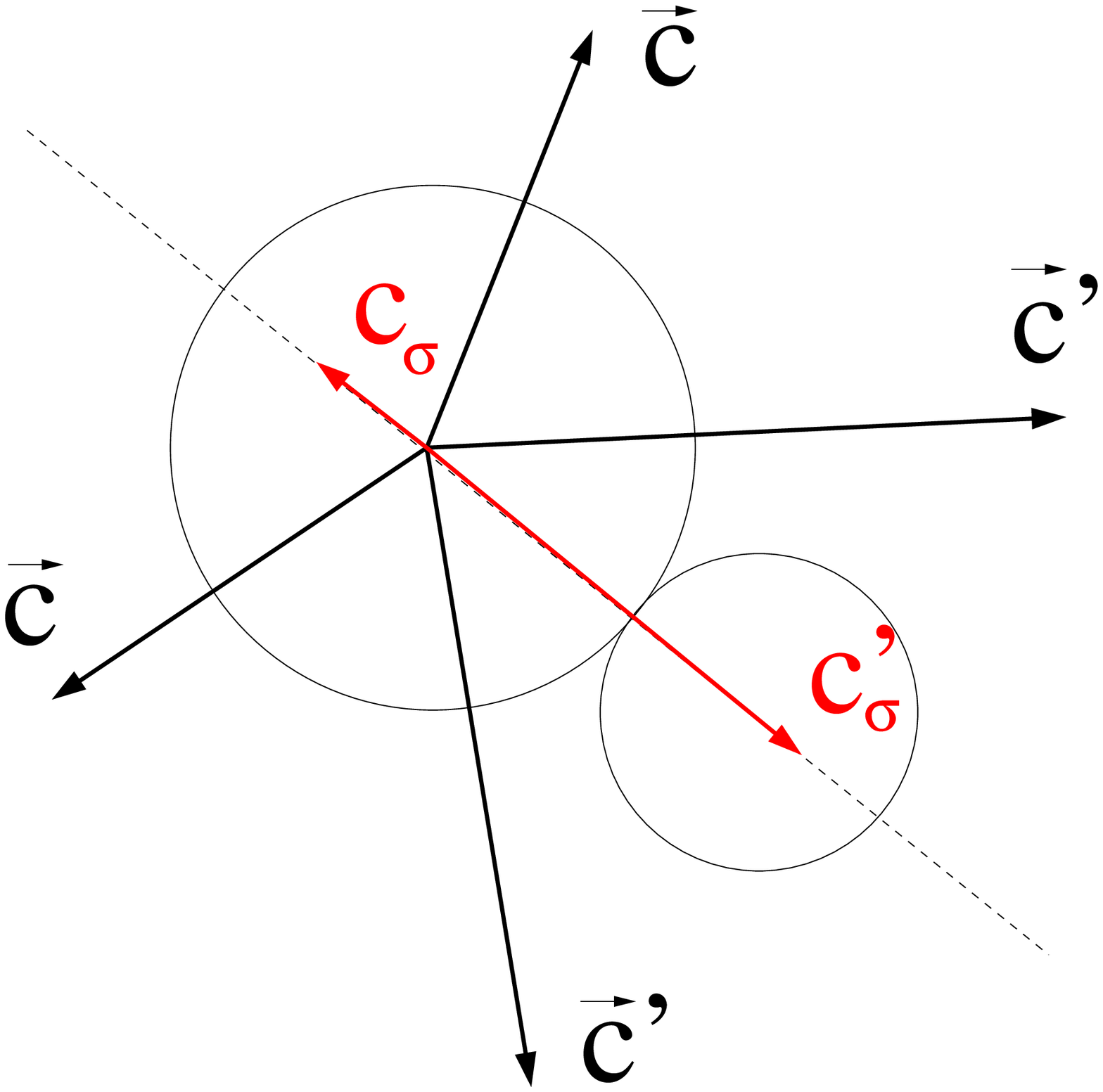}
\caption{An example for the change of variables $(c_x',c_y') \to
  (c_\sigma,c_\sigma')$, introduced in Eq.~\eqref{changevar}. Such
  change of variable, when inverted, has two possible determinations:
  in this example both represented vectors ${\bf c}'$ yield the same
  $(c_\sigma,c_\sigma')$.  \label{fig:cambio}}
\end{figure}
In order to perform the integral, we make the following change of variables (see Fig.~\ref{fig:cambio} for an example)
\begin{eqnarray}
c_{\sigma}&=&c_x\frac{c'_x-c_x}{\sqrt{(c'_x-c_x)^2+(c'_y-c_y)^2}}+
c_y\frac{c'_y-c_y}{\sqrt{(c'_x-c_x)^2+(c'_y-c_y)^2}} \nonumber \\
c'_{\sigma}&=&c'_x\frac{c'_x-c_x}{\sqrt{(c'_x-c_x)^2+(c'_y-c_y)^2}}+
c'_y\frac{c'_y-c_y}{\sqrt{(c'_x-c_x)^2+(c'_y-c_y)^2}} 
\label{changevar}
\end{eqnarray}
which implies

\begin{equation}
d{\bf c}'=dc'_xdc'_y\to dc_{\sigma}dc'_{\sigma}|J|,
\end{equation}
where 

\begin{equation}
|J|=\frac{|c'_{\sigma}-c_{\sigma}|}{\sqrt{c_x^2+c_y^2-c_{\sigma}^2}}\Theta(c_x^2+c_y^2-c_{\sigma}^2)
\end{equation}
is the Jacobian of the transformation.
The collision rate is then
\begin{equation}
r({\bf V})=\chi\sqrt{\frac{\pi}{2T_g/m}}e^{-\frac{mc^2}{4T_g}}
\left[(c^2+2T_g/m)I_0\left(\frac{m c^2}{4T_g}\right)
+c^2 I_1\left(\frac{m c^2}{4T_g}\right)\right],
\end{equation}
where $I_{n}(x)$ are the modified Bessel functions.
For $D_i^{(1)}$ we can write
\begin{eqnarray}
D_i^{(1)}({\bf V})&=&\chi\int_{-\infty}^{+\infty} dc_{\sigma}\int_{c_{\sigma}}^{\infty}dc'_{\sigma}(c_i'-c_i)|J|
\frac{1}{\sqrt{2\pi T_g/m}k(\epsilon)^2} \nonumber \\
&\times&\exp\left\{-m\left[c'_{\sigma}+(k(\epsilon)-1)c_{\sigma}\right]^2/(2T_gk(\epsilon)^2)\right\} \nonumber \\
&=&\chi\int_{-c}^{+c} dc_{\sigma}\int_{c_{\sigma}}^{\infty}dc'_{\sigma}(c_i'-c_i)
\frac{c'_{\sigma}-c_{\sigma}}{\sqrt{c^2-c_{\sigma}^2}} \nonumber \\
&\times&\frac{1}{\sqrt{2\pi T_g/m}k(\epsilon)^2}
\exp\left\{-m\left[c'_{\sigma}+(k(\epsilon)-1)c_{\sigma}\right]^2/(2T_gk(\epsilon)^2)\right\} 
\label{D1}
\end{eqnarray}
where we have enforced the constraint of the theta function,
namely $c_{\sigma}\in(-c,+c)$, with $c=\sqrt{c_x^2+c_y^2}$. Notice that the integral in $dc_{\sigma}'$ is
lower bounded by the condition $c'_{\sigma}\ge c_{\sigma}$ which follows from the definition of $c_{\sigma}$.
In order to compute the integral, we have to invert the transformation~(\ref{changevar}). That yields two
determinations for the variables $c'_x$ and $c'_y$ (see Fig.~\ref{fig:cambio})

\begin{displaymath}
(A) \left\{ \begin{array}{l}
c'_x-c_x =\frac{c'_{\sigma}-c_{\sigma}}{c^2}\left(c_{\sigma}c_x + c_y Sign(c_x)
\sqrt{c^2-c_{\sigma}^2}\right)\\
c'_y-c_y =\frac{c'_{\sigma}-c_{\sigma}}{c^2}\left(c_{\sigma}c_y - c_x Sign(c_x)
\sqrt{c^2-c_{\sigma}^2}\right)\\
\end{array} \right.
\end{displaymath}

\begin{displaymath}
(B) \left\{ \begin{array}{l}
c'_x-c_x =\frac{c'_{\sigma}-c_{\sigma}}{c^2}\left(c_{\sigma}c_x - c_y Sign(c_x)
\sqrt{c^2-c_{\sigma}^2}\right)\\
c'_y-c_y =\frac{c'_{\sigma}-c_{\sigma}}{c^2}\left(c_{\sigma}c_y + c_x Sign(c_x)
\sqrt{c^2-c_{\sigma}^2}\right)
\end{array} \right.
\end{displaymath}
Then the integral~(\ref{D1}) can be written as 

\begin{eqnarray}
D_x^{(1)}({\bf V})&=&\frac{1}{l_0}\int_{-c}^{c} dc_{\sigma}\int_{c_{\sigma}}^{\infty}dc'_{\sigma}
\left[(c_x'-c_x)^{(A)}+(c_x'-c_x)^{(B)}\right]|J| \nonumber \\
&\times&\frac{1}{\sqrt{2\pi T_g/m}k(\epsilon)^2}
\exp\left\{-m\left[c'_{\sigma}+(k(\epsilon)-1)c_{\sigma}\right]^2/(2T_gk(\epsilon)^2)\right\}, \nonumber \\
\end{eqnarray}
yielding

\begin{eqnarray}
D^{(1)}_x&=&-\frac{2}{3}\frac{1}{l_0}k(\epsilon)\sqrt{\frac{m\pi}{2 T_g}}c_x e^{-\frac{m c^2}{4T_g}}\left[
(c^2+3T_g/m)I_0(\frac{m c^2}{4T_g})+(c^2+T_g/m)I_1(\frac{m c^2}{4T_g})\right], \nonumber \\
D^{(1)}_y&=& -\frac{2}{3}\frac{1}{l_0}k(\epsilon)\sqrt{\frac{m\pi}{2 T_g}}c_y e^{-\frac{m c^2}{4T_g}}\left[
(c^2+3T_g/m)I_0(\frac{m c^2}{4T_g})+(c^2+T_g/m)I_1(\frac{m c^2}{4T_g})\right]. \nonumber \\
\end{eqnarray}
Analogously, for the coefficients $D_{ij}^{(2)}$ one obtains

\begin{eqnarray}
D_{xx}^{(2)}({\bf V})&=&\frac{1}{2}\frac{1}{l_0}\int_{-c}^{c} dc_{\sigma}\int_{c_{\sigma}}^{\infty}dc'_{\sigma}
\left[\left((c_x'-c_x)^{(A)}\right)^2+\left((c_x'-c_x)^{(B)}\right)^2\right]|J| \nonumber \\
&\times&\frac{1}{\sqrt{2\pi T_g/m}k(\epsilon)^2}
\exp\left\{-m\left[c'_{\sigma}+(k(\epsilon)-1)c_{\sigma}\right]^2/(2T_gk(\epsilon)^2)\right\} 
\nonumber \\
&=&\frac{1}{2}\frac{1}{l_0}\frac{k(\epsilon)^2}{15}\sqrt{\frac{2m\pi}{T_g}}e^{-\frac{m c^2}{4T_g}} \nonumber \\ 
&\times&\Big\{\Big[c^2(4c_x^2+c_y^2)+3T_g(7c_x^2+3c_y^2)/m+15T_g^2/m^2\Big]I_0\left(\frac{m c^2}{4T_g}\right) \nonumber \\
&+&\Big[c^2(4c_x^2+c_y^2)+T_g(13c_x^2+7c_y^2)/m+3T_g^2/m^2\frac{-c_x^2+c_y^2}{c^2}\Big]I_1\left(\frac{m c^2}{4T_g}\right)\Big\},
\nonumber \\
\end{eqnarray}

\begin{eqnarray}
D_{xy}^{(2)}({\bf V})&=&\frac{1}{2}\frac{1}{l_0}\int_{-c}^{c} dc_{\sigma}\int_{c_{\sigma}}^{\infty}dc'_{\sigma}
\left[(c_x'-c_x)^{(A)}(c_y'-c_y)^{(A)}+(c_x'-c_x)^{(B)}(c_y'-c_y)^{(B)}\right]|J| \nonumber \\
&\times&\frac{1}{\sqrt{2\pi T_g/m}k(\epsilon)^2}
\exp\left\{-m\left[c'_{\sigma}+(k(\epsilon)-1)c_{\sigma}\right]^2/(2T_gk(\epsilon)^2)\right\} 
\nonumber \\
&=&\frac{1}{2}\frac{1}{l_0}\frac{k(\epsilon)^2}{5}\sqrt{\frac{2m\pi}{T_g}}e^{-\frac{m c^2}{4T_g}}c_xc_y \nonumber \\
&\times&\left[(c^2+4T_g/m)I_0\left(\frac{m c^2}{4T_g}\right)+ 
\frac{c^4+2c^2T_g/m-2T_g^2/m^2}{c^2}I_1\left(\frac{m c^2}{4T_g}\right)\right].
\nonumber \\
\end{eqnarray}
Then we introduce the rescaled variables
\begin{equation}
q_x=\frac{c_x}{\sqrt{T_g/m}}\epsilon^{-1} \qquad q_y=\frac{c_y}{\sqrt{T_g/m}}\epsilon^{-1},
\label{rescaled}
\end{equation}
obtaining
\begin{eqnarray}
D_x^{(1)}({\bf V})&=&
-\frac{2}{3}\frac{1}{l_0}\sqrt{\frac{\pi}{2}}\frac{T_g}{m}q_x k(\epsilon)
\epsilon e^{-\frac{\epsilon^2 q^2}{4}}\left[\left(\epsilon^2 q^2+3\right)I_0(\frac{\epsilon^2 q^2}{4})+\left(\epsilon^2 q^2+1\right)
I_1(\frac{\epsilon^2 q^2}{4})\right],\nonumber \\
D_{xx}^{(2)}({\bf V})&=&\frac{1}{2}\frac{1}{l_0}\frac{1}{15}
\sqrt{2\pi}\left(\frac{T_g}{m}\right)^{3/2}k(\epsilon)^2
e^{-\frac{\epsilon^2 q^2}{4}} \nonumber \\ 
&\times&\Big\{\left[\epsilon^4 q^2(4 q_x^2+q_y^2)+3\epsilon^2(7q_x^2+3q_y^2)+15\right]
I_0\left(\frac{\epsilon^2 q^2}{4}\right) \nonumber \\
&+&\left[\epsilon^4 q^2(4q_x^2+q_y^2)+\epsilon^2(13q_x^2+7q_y^2)
+3\frac{-q_x^2+q_y^2}{q^2}\right]I_1\left(\frac{\epsilon^2 q^2}{4}\right)\Big\}  \nonumber \\
D_{xy}^{(2)}({\bf V})&=&\frac{1}{2}\frac{1}{l_0}\frac{1}{5}
\sqrt{2 \pi}\left(\frac{T_g}{m}\right)^{3/2}q_x q_y
k(\epsilon)^2\epsilon^2 e^{-\frac{\epsilon^2 q^2}{4}}
\nonumber \\
&\times&\left[\left(\epsilon^2 q^2+4\right)
I_0\left(\frac{\epsilon^2 q^2}{4}\right) 
+\left(\frac{\epsilon^4 q^4+2\epsilon^2 q^2-2}{\epsilon^2 q^2}\right)I_1\left(\frac{\epsilon^2 q^2}{4}\right)\right].
\label{results}
\end{eqnarray}
Up to this last results we have not introduced any small $\epsilon$
approximation. The next step consists in assuming that $q\sim {\cal
  O}(1)$ with respect to $\epsilon$, which is equivalent to assume
that $c^2\sim T_g/M$: this assumption must be compared to its
consequences, in particular to Eq.~\eqref{Ttracer}; the assumption is
good for not too small values of $\alpha$ and for $\gamma_g \gg
\gamma_b$, i.e. when $T_{tr} \sim T_g$. When this is the case,
expanding in $\epsilon$ and using that $I_0(x)\sim 1+x^2/4$ and
$I_1(x)\sim x/2$ for small $x$, one finds Eqs.~(\ref{largemass}).

\section*{References}

\bibliographystyle{unsrt}
\bibliography{fluct.bib}

\end{document}